\documentclass[sigconf]{acmart}
\AtBeginDocument{%
  }

\copyrightyear{2026}
\acmYear{2026}
\setcopyright{cc}
\setcctype{by}
\acmConference[CIKM '26] {Proceedings of the 35th ACM International Conference on Information and Knowledge Management}{November 7--11, 2026}{Rome, Italy.}
\acmBooktitle{Proceedings of the 35th ACM International Conference on Information and Knowledge Management (CIKM '26), November 7--11, 2026, Rome, Italy}
\acmISBN{979-8-4007-2539-5/2026/11}
\acmDOI{10.1145/3799682.3840709}

\acmSubmissionID{cfp0161}

\usepackage{multirow}
\usepackage{algorithm}
\usepackage{algpseudocode}
\usepackage{balance}
\definecolor{green}{rgb}{0, 0.5, 0}
\definecolor{orange}{rgb}{0.8, 0.6, 0.2}
\definecolor{orange2}{rgb}{1.0, 0.6, 0.2}
\definecolor{red}{rgb}{1.0, 0.0, 0.0}
\definecolor{teal}{rgb}{0.0, 0.4, 0.4}
\definecolor{purple}{rgb}{0.65,0,0.65}
\definecolor{saffron}{rgb}{0.95,0.75,0.2}
\definecolor{turquoise}{rgb}{0.0,0.5,0.5}
\definecolor{black}{rgb}{0.0, 0.0, 0.0}
\definecolor{gray}{rgb}{0.5, 0.5, 0.5}

\newcommand{\ying}[1]{{\color{black}#1}}
\newcommand{\needmodify}[1]{{\color{black}#1}}
\newcommand{\redmarker}[1]{{\color{black}#1}}

\begin{document}

\title{RoE-FND: Synergizing LLMs with Experiential Learning for Effective and Generalizable Evidence-Based Fake News Detection}

\author{Yuzhou Yang}
\email{22110240074@m.fudan.edu.cn}
\affiliation{%
  \institution{Fudan University}
  \city{Shanghai}
  \country{China}
}
\orcid{0000-0001-6957-7682}

\author{Qichao Ying}
\email{shinydotcom@163.com}
\affiliation{%
  \institution{Fudan University}
  \city{Shanghai}
  \country{China}
}
\orcid{0000-0002-6527-2424}

\author{Sheng Li}
\email{lisheng@fudan.edu.cn}
\affiliation{%
  \institution{Fudan University}
  \city{Shanghai}
  \country{China}
}
\orcid{0000-0002-7932-9831}

\author{Zhiying Zhu}
\email{zyzhu@hhu.edu.cn}
\affiliation{%
  \institution{Hohai University}
  \city{Nanjing}
  \country{China}
}
\orcid{0000-0002-1849-3494}

\author{Zhenxing Qian}
\correspondingauthor
\email{zxqian@fudan.edu.cn}
\affiliation{%
  \institution{Fudan University}
  \city{Shanghai}
  \country{China}
}
\orcid{0000-0002-5224-6374}

\author{Xinpeng Zhang}
\email{zhangxinpeng@fudan.edu.cn}
\affiliation{%
  \institution{Fudan University}
  \city{Shanghai}
  \country{China}
}
\orcid{0000-0001-5867-1315}

\renewcommand{\shortauthors}{Yuzhou Yang et al.}

\begin{abstract}
The proliferation of deceptive content in social networks necessitates robust Fake News Detection (FND) systems. Existing pipelines either train detectors on labeled data or leverage Large Language Models (LLMs) for their reasoning ability. However, current approaches remain either limited in generalizability or prone to over-commitment to persuasive yet flawed rationales, lacking systematic experience and mechanisms to expose subtle reasoning errors. 
We propose \textbf{RoE-FND} (\textbf{\underline{R}}eason \textbf{\underline{o}}n \textbf{\underline{E}}xperiences FND), an LLM-based framework that combines self-reflective experience building with deliberation through retrieved experiences for FND. RoE-FND builds an experience bank via reflective learning that compares an unconstrained analysis with a label-conditioned analysis using the ground-truth label as posterior supervision, then summarizes their critical divergence into reusable reasoning guidelines. During inference, RoE-FND generates two opposing deductions via a flipped pseudo-label provided as posterior, retrieves the most relevant experiences for resolving their key disagreement, and adjudicates the better-supported rationale as the final prediction. Experiments across five popular benchmarks, including text-only datasets, i.e., CHEF, Snopes, PolitiFact, and multimedia datasets, i.e., FakeTT, FakeSV, demonstrate that RoE-FND outperforms strong baselines without optimizing LLM parameters on dataset distributions, while exhibiting strong cross-dataset generalization.

\end{abstract}




\maketitle

\section{Introduction}

\redmarker{Online social platforms increasingly distribute news in a multimodal form, where short text is paired with images or videos to attract attention and accelerate sharing. Multimodal fake news refers to information that is intentionally fabricated and can be verified as false, yet is packaged with persuasive cross-modal content that makes it harder for users to judge quickly. Its rapid spread can distort public understanding, amplify polarization, and erode trust in institutions and legitimate media, creating real societal harm beyond simple misinformation. Since manual review is costly and slow at platform scale, automatic Fake News Detection (FND) has become essential for identifying suspicious content early and reducing its downstream impact~\cite{shu2020fakenewsnet}.}

\begin{figure}[!t]
  \centering
  \includegraphics[width=\columnwidth]{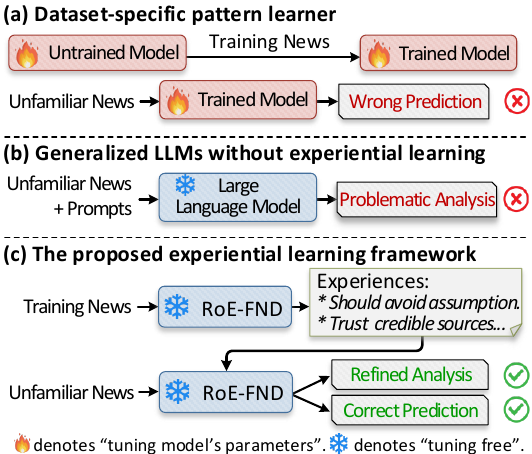}
  \caption{
 Paradigm comparison for multi-modal FND. (a) Training-dependent detectors learn dataset-specific events and patterns, often failing on out-of-domain/unfamiliar news. (b) Generalized LLMs without FND experiential learning avoid dataset bias but can produce unreliable verdicts, e.g., due to hallucinated or flawed rationales. (c) Our proposed RoE-FND addresses both  via synergizing LLMs with experiential learning. We build an experience bank of reusable reasoning guidelines via reflective comparison on past errors, then cross-check different analyses using retrieved task-relevant experiences for final decision.
 }
  \label{fig:intro}
\end{figure}

Existing multi-modal FND schemes can be classified into two categories, namely, content-oriented and evidence-based.
While early content-oriented methods mainly rely on shallow textual features like lexical statistics~\cite{castillo2011information} or syntactic patterns~\cite{feng2012syntactic},
in the past decades, deep-learning-based schemes overwhelm the society, which prepare adequate news samples from the real world, and then develop sophisticated detectors to capture semantic patterns of deception~\cite{shu2020fakenewsnet, popat2017truth,zhang2021mining, zhu2022memory}.
But their decisions are often tied to surface cues and prone to dedicated content editing, and lack explicit grounding mechanism for fact checking.
Evidence-based FND schemes additionally verify news against retrieved evidences~\cite{wu2021unified}, which typically follow a two-stage pipeline as selecting pertinent evidence through lexical similarity~\cite{rashkin2017truth} or neural retrievers~\cite{yang2022coarse}, and learning joint representations of news-evidence pairs to predict~\cite{ma2019sentence, wu2021unified,zhu2022memory}.
\redmarker{However, high-quality logical reasoning can be challenging for traditional neural networks.}

In recent years, Large Language Models (LLMs) have become rising star in both academic and industrial society as they offer strong zero-shot generalization for fake news detection even without requiring task-specific parameter tuning. 
Prior works~\cite{pan2023fact,caramancion2023news} find that when prompted properly, they also exhibit emergent reasoning ability that can be used to analyze claims and synthesize evidence into an explicit rationale. 
Representative methods guide LLMs to execute explicit reasoning programs, incorporate multi-source retrieval to supply structured knowledge, or generate structured justifications to assist downstream decisions~\cite{wei2022emergent,zhu2022memory}.
\ying{
Despite the substantial advancements, there remain certain issues in multi-modal FND. 
Firstly, 
LLMs may over-commit to hallucinated or flawed rationales, rendering the final prediction sometimes doubtful~\cite{huang2023survey,hu2024bad}.
Secondly, existing pipelines provide limited self-correction, so subtle logical flaws in an otherwise persuasive analysis often go unchallenged and thus undetected.
Moreover, although recent debate-based~\cite{han2025debate} and self-reflective~\cite{madaan2023selfreflect} reasoning frameworks improve individual inference trajectories, most existing schemes still treat each case independently and lack a principled mechanism to accumulate reusable verification experiences that can guide future reasoning.
}

\ying{
In this paper, we propose RoE-FND (\underline{\textbf{R}}eason \underline{o}n \underline{\textbf{E}}xperiences FND), an LLM-based framework that treats evidence-based fake news detection as a structured deduction process, requiring the model to produce an explicit, evidence-grounded rationale alongside its verdict.
RoE-FND operates in two stages: an offline self-reflective experience-building stage and an online deliberation stage supported by retrieved experiences, both centered on two analyst agents and an experience bank. 
In the experience-building stage, the Intuitive Analyst first reasons freely about each training case and produces an initial rationale and prediction. If this prediction is incorrect, we then invoke the Informed Analyst, which reasons about the same case under the ground-truth label as posterior, yielding a deliberately contrasted correct analysis. An LLM-based Reflector contrasts the incorrect and correct analyses, distills their critical divergence into compact reasoning guidelines, and stores only those entries that are individually validated to correct the original mistake.
During deployment, RoE-FND elicits two competing analyses with opposite conclusions for each news item. 
An LLM-based Advisor retrieves experiences that encode similar reasoning pitfalls, and an LLM-based Judger evaluates the competing analyses under these guidelines and selects the better-supported one, which directly challenges over-commitment to a single flawed narrative and leverages accumulated reasoning experience to encourage reliability and cross-dataset generalization without dataset-specific fine-tuning.
Although RoE-FND introduces additional inference overhead compared with single-pass prediction pipelines, it prioritizes reasoning reliability, interpretability, and robustness, which are important in the fake news detection task.
The paradigm comparison between existing methods and RoE-FND is shown in Figure~\ref{fig:intro}.
}
We conduct extensive experiments on five challenging datasets, including text-only datasets CHEF~\cite{hu2022chef}, Snopes~\cite{popat2018declare},  PolitiFact~\cite{shu2020fakenewsnet}, and multimedia datasets FakeTT~\cite{fakett}, FakeSV~\cite{fakesv}. 
Results demonstrate that RoE-FND outperforms strong baselines without tuning the applied LLMs, while exhibiting strong cross-dataset generalization.

In summary, the main contributions of this paper are as follows:

\ying{
\begin{itemize}
\item We reframe evidence-based fake news detection as a logical deduction task that requires an explicit, evidence-grounded rationale while avoiding over-commitment to a single persuasive but flawed reasoning. Given a news item, RoE-FND produces two competing deductions with opposite conclusions and selects the better-supported one under guidance from retrieved experiences.
\item We introduce an experiential learning mechanism that distills reusable experiences from the critical divergence between incorrect and correct analyses. Each experience distilled from a training sample is individually verified to improve FND and merged into the experience bank via dynamic experience management.
\item Extensive experiments on both text-only and multimedia benchmarks show strong effectiveness and cross-dataset generalization, and we further analyze experience-bank statistics to explain which experiences are duplicated, decisive, and beneficial after refinement.
\end{itemize}
}

\section{Related Works}
\label{sec:related_works}
\subsection{Multi-modal Fake News Detection}
For content-oriented FND, the general paradigm is to transform the news into a multi-dimensional latent representations and learn binary separators without the help of evidence retrieval. 
While many works uncover unimodal distributional difference of real and fake news such as image composition~\cite{jin2016novel}, manipulation traces~\cite{Leveraging_5}, and stylistic cues~\cite{FND-Survey}, multi-modal content-oriented FND can benefit from multi-modal feature extraction and fusion~\cite{SpotFake,zhou2020mathsf}, or additionally highlight cross-modal correlation~\cite{wei2022cross,WWW}. 
For example, Zhou et al.~\cite{zhou2023multimodal} uses CLIP to analyze text and image embedding similarity, and Ying et al.~\cite{ying2023bootstrapping} bootstraps multi-modal and multi-view representations for decision.
\ying{Nonetheless, without retrieval capability for factual grounding, the model must implicitly ``remember'' event-specific patterns from training data, making it more prone to overfitting to particular events, dataset distributions.}

For evidence-based FND, the general paradigm is to conduct knowledge comparison between news and relevant evidence materials~\cite{zhou2020survey}. 
For example, DeClarE~\cite{popat2018declare} proposes the earliest evidence-based FND method, it jointly learns representations of news content with evidence materials.
Many works leverage the merits of hierarchical attention for evidence-news interaction, e.g., HAN~\cite{ma2019sentence}, EHIAN~\cite{wu2021evidence}, MAC~\cite{vo2021hierarchical}.
GET proposes a graph neural network to model distant semantic correlation among news and evidence materials~\cite{xu2022evidence}.
MUSER retrieves the key evidence information for news verification in a multi-round retrieval process~\cite{liao2023muser}.
SEE builds a dynamic network to adaptively exploit the coarse evidence materials~\cite{yang2024see}.
\ying{Despite their effectiveness, existing evidence-based detectors often lack an explicit mechanism to verify and justify the relevance of used materials and may still provide limited, hard-to-audit decision logic, which can hinder robustness and trustworthy deployment on unfamiliar domains.}

\subsection{LLM-aided Fake News Detection}
\redmarker{Many agentic workflow for multimodal FND are proposed in recent years, significantly improving the overall accuracy of detection and interpretability of the results.} 
For example, ProgramFC~\cite{pan2023fact} is a program-guided framework that verifies news by generating and executing explicit reasoning programs with large language models.
IMRRF~\cite{li2025imrrf} improves detection performance by retrieving evidence from multiple knowledge sources and transforming it into structured world knowledge for LLM-based reasoning.
L-Defense~\cite{wang2024explainable} extracts competing evidences and generates structured justifications using large language models, followed by training classifiers to predict the final label.
SheepDog~\cite{wu2024sheep} employs LLM-generated, style-diverse reframings to enable style-agnostic training, encouraging detectors to focus on news content rather than writing style.
Nevertheless, recent LLM reasoning frameworks have increasingly explored debating~\cite{han2025debate}, self-reflection~\cite{madaan2023selfreflect} to improve reasoning reliability by exposing conflicting viewpoints and revising flawed rationales. 
Despite these advances, existing LLM-aided fake news detection pipelines generally lack a systematic mechanism to accumulate and reuse transferable reasoning experiences distilled from past failures, limiting their ability to stabilize future verification on unfamiliar cases and domains.


\begin{figure*}[!t]
  \centering
  \includegraphics[width=1.0\textwidth]{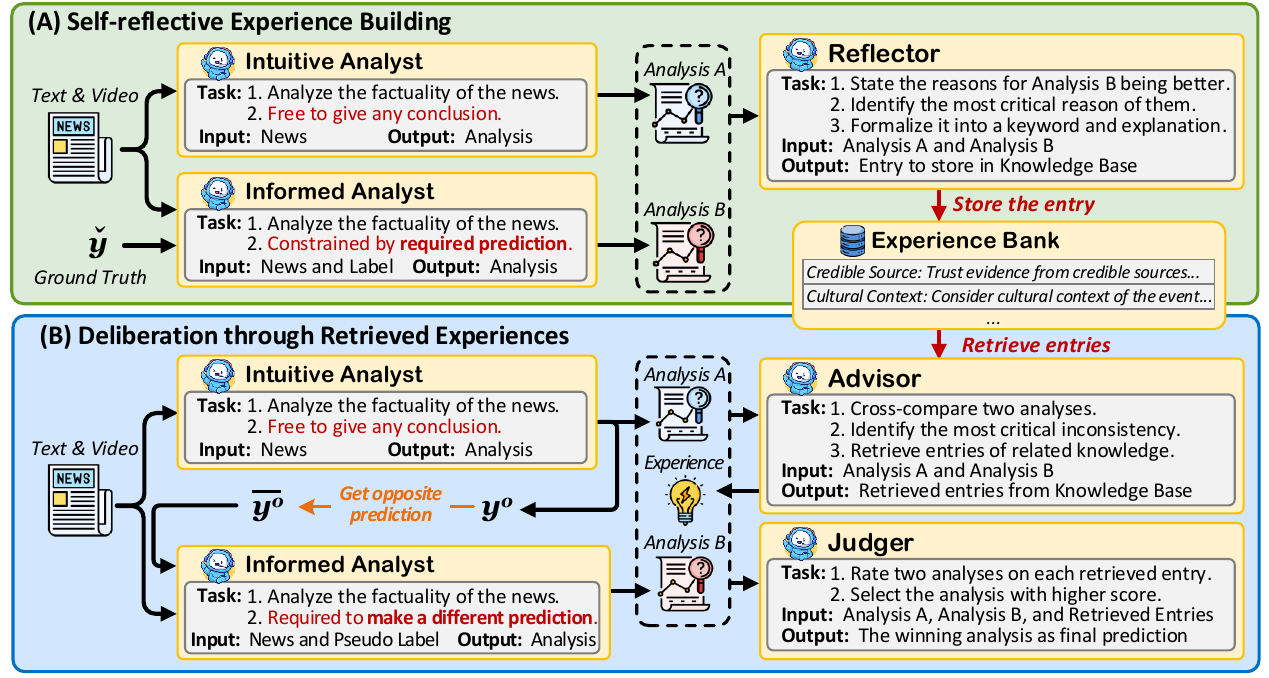}
  \caption{ 
The RoE-FND framework consists of two stages. 
In Stage A, RoE-FND builds up an experience bank through reflection. For each sample input, if it is wrongly predicted, we let the analyst write an analysis by revealing the ground truth ahead. Then, RoE-FND produces experience by cross-comparing the two opposite analyses to identify flaws within the original one. 
In Stage B, RoE-FND produces two analyses with opposite conclusions. Then it retrieves experience accordingly from the built bank to assist in selecting the better analysis as the final prediction. 
}
  \label{fig:architecture}
\end{figure*}

\section{Proposed Method}

\ying{Figure~\ref{fig:architecture} depicts the pipeline of the proposed Reason on Experiences (RoE) fake news detection approach, which consists of two stages, i.e., the offline self-reflective experience-building stage, and the real-world application stage for online prediction via deliberation through retrieved experiences. 
In both stages, RoE-FND employs two agent-based \texttt{Analyst}s together with an experience bank, i.e., the \texttt{IntuitiveAnalyst} and the \texttt{InformedAnalyst}. The \texttt{IntuitiveAnalyst} examines the news with retrieved evidence and produces a free-form rationale and verdict. The \texttt{IntuitiveAnalyst} generates a rationale under a required posterior prediction. In the experience-building stage, we only invoke the \texttt{InformedAnalyst} when the \texttt{IntuitiveAnalyst} makes an incorrect prediction, so their rationales provide a deliberately contrasted view of the same case. Besides, we employ a \texttt{Reflector} in the first stage, to condense critical differences between a correct and an incorrect analysis into a reusable experience entry and maintains the experience bank. In the second stage, there is a \texttt{Advisor} that inspects a pair of current analyses, pinpoints their main inconsistency, and retrieves the most relevant experiences to guide the decision. 
Finally, there is a \texttt{Judger} that evaluates the competing analyses under this retrieved guidance and outputs the final prediction and corresponding reasoning.
}

\subsection{Self-reflective Experience Building}

In this stage, we structure the workflow of the RoE-FND to produce experiences by exploring samples in the training set.

\noindent\textbf{\ying{Reflective Dual Analysis.}}
We utilize the \texttt{IntuitiveAnalyst} and the \texttt{InformedAnalyst} to generate dual analysis with reflective value.
Both \texttt{Analyst} agents are equipped with an internet searching tool, which enables them to retrieve factual information by querying search engines for their reasoning.
Specifically, the \texttt{IntuitiveAnalyst} is prompted in the chain-of-thought (CoT) style~\cite{wei2022chain}, requiring it analyze the news's factuality step-by-step, and finally outputs an analysis with a prediction $y\in\left\{true, false\right\}$.
\ying{If the prediction is correct, we further rerun the analysis randomly within two rounds to see if the predictions can be consistently correct. 
No error prediction indicates no reflective experience from the current news, Otherwise, e.g., due to LLM's internal sampling, we}
prompt the \texttt{InformedAnalyst} also in chain-of-verification style~\cite{dhuliawala2024chain}, which reveals the ground truth prediction to it ahead so that the analysis can be more reliable, i.e.,
\begin{align}
\begin{split}
y, {\mathcal{A}} &= \texttt{IntuitivelAnalyst}\left(\mathcal{N}_{train}\right), \\
\tilde{y}, {\overline{\mathcal{A}}} &= \texttt{InformedAnalyst}\left(\mathcal{N}_{train}, \check{y}\right), \\
\textbf{s.t. }& \check{y} =\tilde{y} \neq y,
\end{split}
\label{eq:dual_channel}
\end{align}
where $\check{y}\in\{true, false\}$ denotes the ground truth label of the news $\mathcal{N}$, and we regulate that $\tilde{y} = \check{y}$.
The dual analyzing produces paired contradictory analyses for reflection.



\noindent\textbf{Experience Generation via Comparative Reflection.}
We use the \texttt{Reflector} for experience generation.
Instead of directly asking it to examine the problematic content~\cite{asaiself, jeong2024adaptive}, we mark the correct analysis and let it cross-compare $\mathcal{A}$ and $\overline{\mathcal{A}}$ for reflection.
Specifically, we prompt it in the CoT style to perform three sub-tasks. 
To begin with, it is asked to reason why $\overline{\mathcal{A}}$ is correct while $\mathcal{A}$ is wrong.
Then, it is asked to identify the most critical reason among them.
Third, it is asked to produce an experience by formatting the reason into a keyword and a short explanation \textit{without mentioning any entity}.
This design discourages event-oriented experiences and promotes generalizable reasoning guidelines. In other words, the key is not which evidence is cited in a specific instance, but what reusable verification experience can be distilled from the evidence–rationale discrepancy.
For each newly produced experience, we also enforce a safeguard to verify that injecting it into the prompt of \texttt{IntuitiveAnalyst} can correct the previously wrong prediction before committing it to the bank, i.e.,
\begin{align}
\begin{split}
&\mathcal{R} = \texttt{Reflector}\left(\mathcal{A}, {\overline{\mathcal{A}}}\right),\\
& \textbf{s.t. } \check{y}=\texttt{IntuitiveAnalyst}(\mathcal{N},\mathcal{R}),
\end{split}
\end{align}
where $\mathcal{R}$ denotes derived experience. 
Note that we refrain from finetuning the \texttt{Analyst} and solely mine $\mathcal{R}$ that could help the naive LLM make individual correct predictions. 
Unlike debate-based reasoning frameworks that iteratively exchange arguments to defend competing conclusions, RoE-FND uses dual analyses primarily as reflective supervision signals for experience construction. 
Our objective is to identify the critical reasoning divergence that explains why one rationale succeeds while another fails. 
This design enables the framework to distill compact and reusable reasoning guidance without generating increasingly lengthy debate trajectories.

\noindent\textbf{\needmodify{Dynamic Experience Management.}}
\redmarker{The experience bank, denoted as $\mathcal{B}$, cannot grow indefinitely and must be efficient, explainable and inspiring.} 
Thus, we design an experience management strategy that serves two primary objectives, i.e., \redmarker{common experience refinement and extreme experience elimination.}
First, it compresses redundant experience entries.
Since reflection can generate similar experiences across different samples, we prevent duplication by evaluating semantic similarity between the experience candidate and those within the bank.
When the \texttt{Reflector} creates a new entry, we compute the semantic similarity between the keywords of the new entry and those already in storage. The entry is saved only if the maximum similarity is below a pre-defined threshold, otherwise, we ask LLM to merge two similar entries.
Second, we refine the bank by filtering out extreme experiences, i.e., those that apply only to narrow cases and lack generalizability.
We propose average advantage score, denoted as $Adv$, for entries within the unrefined bank $\mathcal{B}^*$, which are calculated through the \texttt{Judger}.
Specifically, it scores the correct and wrong analyses w.r.t each retrieved entry within 1 to 10. Suppose the $k$-th entry is retrieved, we thereby get its score as $s_k=\max({s_k^{\text{correct}}-s_k^{\text{wrong}}},0)$, i.e.,
\begin{align}
    &{Adv}_{k}=\frac{1}{N_k}\sum^{N_k}_{i=1} s^i_k, \\
    & \mathcal{B} = \left\{R_k|Adv_k\ge th_{dis},R_k\in\mathcal{B}^*\right\},
\end{align}
where we empirically set the threshold ${th}_{dis}$ to discard entries with advantages below it.
This procedure preserve effective entries to improves the quality of the experience bank.

\begin{algorithm}[t]
\caption{Stage I: Self-reflective Experience Building}
\label{alg:experience_building}
\begin{algorithmic}[1]
\Require Training set $\mathcal{D}_{train}$, initial experience bank $\mathcal{B}=\emptyset$, similarity threshold $\tau_s$, discard threshold $\tau_d$
\Ensure Refined experience bank $\mathcal{B}$

\For{each training news sample $(\mathcal{N}_i, \check{y}_i) \in \mathcal{D}_{train}$}
    \State $(y_i, A_i) \gets \textsc{IntuitiveAnalyst}(\mathcal{N}_i)$

    \If{$y_i = \check{y}_i$}
        \State Re-run $\textsc{IntuitiveAnalyst}(\mathcal{N}_i)$ for consistency check
        \If{all predictions remain correct}
            \State \textbf{continue}
        \EndIf
    \EndIf

    \State $(\tilde{y}_i, \tilde{A}_i) \gets \textsc{InformedAnalyst}(\mathcal{N}_i, \check{y}_i)$
    \State \textbf{constrain} $\tilde{y}_i = \check{y}_i \neq y_i$

    \State $R_i \gets \textsc{Reflector}(A_i, \tilde{A}_i)$
    \Comment{Turn key divergence into reusable experience}

    \State $(y_i^{R}, A_i^{R}) \gets \textsc{IntuitiveAnalyst}(\mathcal{N}_i, R_i)$

    \If{$y_i^{R} = \check{y}_i$}
        \State $R_j \gets \arg\max_{R \in \mathcal{B}} \operatorname{sim}(R_i, R)$

        \If{$\operatorname{sim}(R_i, R_j) < \tau_s$}
            \State $\mathcal{B} \gets \mathcal{B} \cup \{R_i\}$
        \Else
            \State $R_j \gets \textsc{MergeExperience}(R_i, R_j)$
        \EndIf
    \EndIf
\EndFor

\For{each experience entry $R_k \in \mathcal{B}$}
    \State Compute its average advantage score:
    \[
    Adv_k =
    \frac{1}{N_k}
    \sum_{i=1}^{N_k}
    \max \left(
    s^{correct}_{i,k} - s^{wrong}_{i,k}, 0
    \right)
    \]
\EndFor

\State $\mathcal{B} \gets \{R_k \mid Adv_k \geq \tau_d,\ R_k \in \mathcal{B}\}$

\Return $\mathcal{B}$
\end{algorithmic}
\end{algorithm}

\subsection{Deliberation through Retrieved Experiences}

In the inference stage, we leverage the distilled experience bank to guide deliberation between competing analyses.

\noindent\textbf{\ying{Deliberative Dual Analysis.}}
\ying{We again adopt dual analysis instead of a single-shot prediction, explicitly constructing two analyses that support contradictory conclusions. This design aligns with the experience bank, whose entries are distilled from key divergences between correct and incorrect reasoning, and allows us to leverage these experiences when cross-comparing the analyses to identify the more reliable verdict.}
Concretely, we first obtain a free-form analysis $\mathcal{A}$ and prediction $y^o$ by the \texttt{IntuitiveAnalyst}.
\ying{To minimize the impact from LLM random sampling, here we also infer the model for three rounds and choose one with the overwhelming binary prediction.}
Then, we provide the \texttt{InformedAnalyst} with the opposite pseudo label $\overline{y^o}$, as posterior, i.e.,
\begin{align}
\begin{split}
y^o,\mathcal{A}^o&=\texttt{IntuitiveAnalyst}(\mathcal{N}_{test}), \\
\tilde{y^o},\overline{\mathcal{A}^o}&=\texttt{InformedAnalyst}(\mathcal{N}_{test},\overline{y^o}), \\
\textbf{s.t. }& \tilde{y^o} = \overline{y^o}\neq y^o.
\end{split}
\end{align}
This produces two opposing analyses for downstream deliberation.

\noindent\textbf{\needmodify{Experience Retrieval and Integration}.}
\ying{The \texttt{Advisor} retrieves entries from the experience bank according to the news content and the provided analyses.}
We devise a prompt in CoT style to complete the mission by three steps.
First, it is asked to cross-compare two analyses and state their discrepancies.
Then, it is asked to identify the most critical point that bifurcates the two analyses.
Lastly, it is asked to summarize the point into a sentence $Q$ as the query to retrieve relevant experiences. 
E.g., if one analysis adopts an opinion that most evidence supports, while the other analysis uses the opposite opinion from a credible source, it may generate ``whether use the majority evidence or trust credible sources''.
It retrieves top-$k$ relevant entries based on sentence-wise semantic similarity calculation. 
We set a threshold ${th}_{re}$ to ensure the relevance, i.e., if the maximum of similarity scores is below ${th}_{re}$, we initiate a re-retrieval process by prompting \texttt{Advisor} to refine the query.
Let the retrieved experiences be $\mathcal{C}$, we have:
\begin{align}
\begin{split}
\mathcal{Q} &= \texttt{Advisor}\left(\mathcal{A},\overline{\mathcal{A}}\right), \\
\mathcal{C} &= [R_1,\dots,R_k] = \arg\max_{\mathcal{R}_x\in \mathcal{B}}{{\operatorname{sim}(\mathcal{R}_x,\mathcal{Q})}},\\
&\textbf{s.t. } \min_{R_i\in\mathcal{C}}{\operatorname{sim}(R_i, Q)} \ge th_{re}.
\end{split}
\end{align}


\noindent\textbf{Prediction via Deliberation.}
In the last step, we assign a \texttt{Judger} to play a role of an expert in rating and assessment.
We devise a prompt with an experience-grounded scorecard that rates both analyses on each retrieved entry with short evidence-based justifications and includes a label–rationale consistency check before selecting the final verdict.
The conclusion of the winning analysis is regarded as the framework's final prediction.
\texttt{Judger}'s procedure to output the final prediction is represented as:
\begin{align}
\begin{split}
\mathcal{J} &= \texttt{Judger}\left(\mathcal{A}^o, \overline{\mathcal{A}^o},\mathcal{C}\right),\\
y&= \begin{cases}
{y}^o, \emph{\space\space if \space} \mathcal{J} \emph{\space indicates\space} \mathcal{A^o}\geq \overline{\mathcal{A}^o}, \\
\overline{y^o}, \emph{\space\space otherwise},
\end{cases}
\end{split}
\end{align}
where we utilize the symbol $\ge$ to represent favoring $\mathcal{A}^o$ over $\overline{\mathcal{A}^o}$.

\begin{algorithm}[t]
\caption{Stage II: Deliberation through Retrieved Experiences}
\label{alg:deliberation}
\begin{algorithmic}[1]
\Require Test news sample $\mathcal{N}$, refined experience bank $\mathcal{B}$, top-$k$, retrieval threshold $\tau_r$
\Ensure Final prediction $\hat{y}$ and final rationale $\hat{A}$

\State $(y^{o}, A^{o}) \gets \textsc{IntuitiveAnalyst}(\mathcal{N})$
\Comment{Free-form analysis}

\State $\bar{y}^{o} \gets \textsc{OppositeLabel}(y^{o})$

\State $(\tilde{y}^{o}, \tilde{A}^{o}) \gets \textsc{InformedAnalyst}(\mathcal{N}, \bar{y}^{o})$
\State \textbf{constrain} $\tilde{y}^{o} = \bar{y}^{o} \neq y^{o}$

\State $Q \gets \textsc{Advisor}(A^{o}, \tilde{A}^{o})$
\Comment{Identify key inconsistency and form retrieval query}

\State $\mathcal{C} \gets \textsc{RetrieveTopK}(\mathcal{B}, Q, k)$

\While{$\max_{R_i \in \mathcal{C}} \operatorname{sim}(R_i, Q) < \tau_r$}
    \State $Q \gets \textsc{AdvisorRefineQuery}(A^{o}, \tilde{A}^{o}, Q)$
    \State $\mathcal{C} \gets \textsc{RetrieveTopK}(\mathcal{B}, Q, k)$
\EndWhile

\State $J \gets \textsc{Judger}(A^{o}, \tilde{A}^{o}, \mathcal{C})$
\Comment{Score both analyses under retrieved experiences}

\If{$J$ favors $A^{o}$}
    \State $\hat{y} \gets y^{o}$
    \State $\hat{A} \gets A^{o}$
\Else
    \State $\hat{y} \gets \tilde{y}^{o}$
    \State $\hat{A} \gets \tilde{A}^{o}$
\EndIf

\Return $\hat{y}, \hat{A}$
\end{algorithmic}
\end{algorithm}


\subsection{Implementation Details}
\label{section_imp_detail}
\noindent\textbf{Framework Settings.}
We utilize the OpenAI platform and the DeepSeek platform to run RoE-FND with their models.
Qwen-7B is run locally with vLLM. Invocations of these models use OpenAI SDK standard.
And we utilzie DashScope SDK to invoke Qwen-VL from Aliyun platform, which allows to upload videos for processing. 
The temperatures of LLMs are set to $1.0$.
The similarity threshold for experience retrieval is empirically set ${th}_{re}=0.9$, and the discarding threshold is empirically set as ${th}_{dis}=0.25$.
For \texttt{Advisor}, it retrieves $k=3$ entries.
\ying{For the \texttt{Analysts}, if the dataset is already with off-line supplementary materials, similar to many evidence-based FND baselines, we downgrade the searching tool to only utilize these prepared materials, otherwise, online searching is granted with restriction to filter out the debunk news or materials with identical or very similar content.}
All modules are inferred with dedicated prompting, without being explicitly fine-tuned with any FND label.
Prompts of LLM-based modules are exhibited in the appendix.

\begin{table*}[!t]
\centering
\caption{Performance comparison based on text-only FND datasets. We report accuracy (ACC), F1-Macro (F1), precision (PR), and recall (RC). * denotes the performance of enhanced RoE-FND via knowledge distillation from DeepSeek R1.}
\setlength{\tabcolsep}{1.4mm}
\begin{tabular}{llcllll|llll|llll}
\toprule
\multirow{2}{*}{\textbf{Baseline Method}} & \multirow{2}{*}{\textbf{Ref.}} & \multirow{2}{*}{\textbf{Trained}} & \multicolumn{4}{c}{\textbf{CHEF}} & \multicolumn{4}{c}{\textbf{Snopes}} & \multicolumn{4}{c}{\textbf{PolitiFact}}\\ 
\cmidrule{4-15}
 & & & ACC & F1 & PR & \multicolumn{1}{l}{RC} & ACC & F1 & PR & \multicolumn{1}{l}{RC} & ACC & F1 & PR & RC \\
\hline
GET & ~\cite{xu2022evidence} & \checkmark & 0.588 & 0.602 & 0.585 & 0.582 & 0.814 & 0.771 & 0.721 & 0.854 & 0.694 & 0.691 & 0.687 & 0.708 \\
ReRead & ~\cite{hu2023read} & \checkmark & 0.789 & 0.776 & 0.826 & 0.745 & 0.816 & 0.714 & 0.652 & 0.789 & 0.693 & 0.681 & 0.711 & 0.718 \\
SheepDog & ~\cite{wu2024sheep} & \checkmark & 0.601 & 0.602 & 0.601 & 0.632 & 0.718 & 0.681 & 0.679 & 0.733 & 0.648 & 0.652 & 0.658 & 0.648 \\
MUSER & ~\cite{liao2023muser} & \checkmark & 0.608 & 0.612 & 0.603 & 0.631 & 0.841 & 0.745 & 0.699 & 0.798 & 0.729 & {{0.732}} & 0.735 & 0.728 \\
SEE & ~\cite{yang2024see} & \checkmark & 0.763 & 0.776 & 0.751 & 0.802 & 0.824 & 0.786 & {{0.773}} & 0.845 & 0.706 & 0.705 & 0.688 & 0.724 \\
L-Defense & ~\cite{wang2024explainable} & \checkmark & 0.711 & 0.708 & 0.692 & 0.721 & 0.773 & 0.761 & 0.755 & 0.792 & 0.695 & 0.691 & 0.684 & 0.711 \\
\hline
LLaMa-7B & ~\cite{touvron2023llama2} & - & 0.646 & 0.662 & 0.836 & 0.469 & 0.703 & 0.431 & 0.487 & 0.386 & 0.555 & 0.416 & 0.613 & 0.315 \\
Qwen-7B & ~\cite{yang2024qwen2} & - & 0.622 & 0.510 & 0.615 & 0.506 & 0.671 & 0.235 & 0.362 & 0.174 & 0.519 & 0.439 & 0.597 & 0.141 \\
GPT-4o-mini & ~\cite{openai2022chatgpt} & - & 0.740 & 0.773 & 0.687 & 0.883 & 0.745 & 0.452 & 0.604 & 0.361 & 0.575 & 0.695 & 0.648 & 0.775 \\
DeepSeekv3 & ~\cite{liu2024deepseek} & - & 0.780 & 0.784 & 0.790 & 0.778 & 0.736 & 0.253 & 0.712 & 0.154 & 0.530 & 0.631 & 0.485 & {0.872} \\
RAG (DeepSeekv3) & ~\cite{lewis2020retrieval} & - & 0.811 & 0.800 & 0.832 & 0.800 & 0.820 & 0.711 & 0.730 & 0.800 & 0.707 & 0.688 & 0.705 & {0.901} \\
Self-Reflect (DeepSeekv3) & ~\cite{madaan2023selfreflect} & - & 0.789 & 0.785 & 0.790 & 0.780 & 0.814 & 0.815 & 0.821 & 0.810 & 0.709 & 0.704 & 0.704 & {0.704} \\
D2D (DeepSeekv3) & ~\cite{han2025debate} & - & 0.820 & 0.816 & 0.821 & 0.811 & 0.818 & 0.811 & 0.766 & 0.841 & 0.714 & 0.709 & 0.709 & {0.709} \\
ProgramFC & ~\cite{pan2023fact} & - & 0.694 & 0.708 & 0.723 & 0.697 & 0.741 & 0.619 & 0.542 & 0.723 & 0.678 & 0.684 & 0.725 & 0.741 \\
ProgramFC (DeepSeekv3) & ~\cite{pan2023fact}  & - & 0.750 & 0.743 & 0.773 & 0.791 & 0.769 & 0.776 & 0.758 & 0.795 & 0.683 & 0.689 & 0.748 & 0.661 \\
ProgramFC (4o-mini) & ~\cite{pan2023fact} & - & 0.748 & 0.751 & 0.761 & 0.788 & 0.769 & 0.776 & 0.758 & 0.795 & 0.692 & 0.666 & 0.653 & 0.856 \\
IMRRF (DeepSeekv3) & ~\cite{li2025imrrf} & - & 0.799 & 0.802 & 0.811 & 0.801 & 0.798 & 0.781 & 0.772 & 0.818 & 0.693 & 0.681 & 0.654 & 0.737 \\
IMRRF (4o-mini) & ~\cite{li2025imrrf} & - & 0.776 & 0.769 & 0.764 & 0.781 & 0.802 & 0.784 & 0.768 & 0.800 & 0.703 & 0.700 & 0.755 & 0.681 \\
OpenAI o1 & ~\cite{jaech2024o1} & - & 0.890 & 0.888 & 0.875 & 0.892 & 0.852 & \underline{0.800} & \textbf{0.788} & 0.823 & \underline{0.833} & \textbf{0.833} & \underline{0.832} & 0.820 \\
DeepSeek-R1 & ~\cite{guo2025deepseekr1} & - & 0.882 & 0.880 & 0.870 & 0.891 & 0.876 & 0.788 & \underline{0.783} & 0.791 & 0.800 & \underline{0.792} & 0.775 & \textbf{0.899} \\
\hline
RoE-FND (Qwen-7B) &  & - & 0.828 & 0.827 & 0.846 & 0.784 & 0.824 & 0.747 & 0.644 & \underline{0.888} & {{0.743}} & 0.709 & {\textbf{0.875}} & 0.659\\
RoE-FND (DeepSeekv3) &  & - & 0.890 & 0.892 & {{0.876}} & 0.908 & \underline{0.863} & {0.791} & 0.708 & \textbf{0.896} & 0.731 & 0.758 & 0.767 & 0.676 \\
RoE-FND (4o-mini) &  & - & {\underline{0.891}} & {\underline{0.893}} & {\underline{0.876}} & {\underline{0.911}} & 0.860 & 0.774 & 0.732 & 0.822 & 0.732 & {{0.735}} & 0.662 & {{0.827}}  \\
RoE-FND$^*$ (Qwen-7B) &  & \checkmark & {\textbf{0.904}} & {\textbf{0.908}} & {\textbf{0.899}} & {\textbf{0.914}} & \textbf{0.886} & \textbf{0.818} & {0.773} & 0.842 & \textbf{0.891} & {{0.765}} & 0.762 & \underline{0.833}  \\
\hline
\end{tabular}
\label{tab:comparison}
\end{table*}


\noindent\textbf{Efficiency Considerations.}
Although RoE-FND adopts a multi-stage deliberative reasoning pipeline, several mechanisms are introduced to reduce practical inference overhead. The \texttt{Reflector} only participates in offline experience construction and is excluded during deployment-time prediction. Retrieved experiences are compact text rather than lengthy demonstrations, keeping the \texttt{Advisor} and \texttt{Judger} lightweight in token consumption.

To further improve deployment efficiency, we distill reasoning traces from strong teacher models, e.g., DeepSeek-R1, into lightweight student models, e.g., Qwen-7B. Specifically, teacher-generated reasoning traces are first constructed under posterior supervision utilizing the \texttt{InformedAnalyst} and then used to fine-tune lightweight \texttt{Analyst}s through parameter-efficient LoRA~\cite{hu2022lora}. During deployment, the distilled student models replace heavyweight \texttt{Analyst}s, substantially reducing latency while preserving strong reasoning capability. To facilitate reproducibility, we provide detailed prompting templates, module configurations, and implementation settings in the appendix.

\section{Experiments}

\subsection{Experimental Setups}
\noindent\textbf{Datasets.}
\redmarker{We conduct experiments based on famous text-only and multi-modal FND datasets. For text-only datasets,}
CHEF~\cite{hu2022chef} is a Chinese dataset collected from the real world with multiple domains of news. 
PolitiFact~\cite{rashkin2017truth}, and Snopes~\cite{popat2018declare} for our experiments are collected from fact-checking websites.
CHEF contains 5,754 samples for training, 666 samples each for validation and test.
PolitiFact contains 2,140 samples for training, 714 for validation, and 714 for testing.
Snopes contains 2,604 samples for training, 869 each for validation and test.
Notably, all data from these datasets are collected from the real world and are prepared with fixed evidence articles corresponding to each piece of news.
These datasets remain valuable because they provide standardized evidence materials, enabling controlled comparison of reasoning quality across methods.
\redmarker{For multimedia datasets, }
FakeSV~\cite{fakesv} is a Chinese dataset for fake news detection on short video platforms.
FakeTT~\cite{fakett} is an English dataset for fake news detection curated from TikTok.
For all datasets, we utilize the official data splitting either directly from dataset source or specified in baseline works.

\noindent\textbf{Baseline Setups.}
We primarily use performance reported in the original baseline papers if available. 
Otherwise, we run their public code following the same implementation details.
For the LLM-based methods, we apply the same prompts to ensure fair comparison.  
For ProgramFC~\cite{pan2023fact} and IMRRF~\cite{li2025imrrf}, we upgrade the LLM component to a more capable model—DeepSeekv3 and GPT-4o-mini, while the other configurations are kept unchanged.

\begin{table}[!t]
\centering
\caption{Performance comparison based on multimedia (short-video platform) FND datasets.}
\setlength{\tabcolsep}{0.8mm}
{
  \begin{tabular}{lcc|rr|rr}
  \hline
  \multirow{2}{*}{\centering\textbf{Method}} &   \multirow{2}{*}{\centering\textbf{Ref.}} &   \multirow{2}{*}{\centering\textbf{Trained}} & \multicolumn{2}{c|}{\textbf{FakeTT}} & \multicolumn{2}{c}{\textbf{FakeSV}}
  \\
  \cline{4-7} & & & ACC & F1  & ACC & F1 \\ 
\hline
SV-FEND & \cite{fakesv} & \checkmark & 0.771 & 0.756 & 0.808 & 0.805 \\
FakingRecipe & \cite{fakett} & \checkmark & 0.792 & 0.777 & 0.854 & 0.848 \\
ExMRD & \cite{hong2025exmrd} & \checkmark & 0.842 & 0.831 & 0.869 & 0.865 \\
\redmarker{CA-FVD} & \cite{wang2025consistency} & \checkmark & 0.816 & 0.803 & 0.858 & 0.853 \\
Qwen-VL & \cite{qwenvl} & - & 0.659 & 0.631 & 0.641 & 0.641 \\
Qwen-VL + RAG & \cite{qwenvl} & - & 0.710 & 0.742 & 0.694 & 0.707 \\ 
\hline
RoE-FND (Qwen-VL) & & - & \textbf{0.882} & \textbf{0.878} & \textbf{0.890} & \textbf{0.888}\\
    \hline
  \end{tabular}
  }
\label{tab:multimedia}
\end{table}

\begin{table}[!t]
\centering
\caption{Results of cross-datasets testing. The suffix of ACC denotes the detection accuracy of the models trained on Snopes (-S) and PolitiFact (-P). {$\downarrow$}(\%) indicates the decrease compared to testing on the same dataset.}
\setlength{\tabcolsep}{1.2mm}
{
\begin{tabular}{l|rrr|rrr}
\hline
\multirow{2}{*}{\centering\textbf{Method}} & \multicolumn{3}{c|}{\textbf{Snopes}} & \multicolumn{3}{c}{\textbf{PolitiFact}}
  \\
  \cline{2-7}
& {ACC-S} & {ACC-P} & {$\downarrow$}(\%) & {ACC-P} & {ACC-S}  & {$\downarrow$}(\%) \\ 
     \hline
GET & 0.814 & 0.662 & 15.2 & 0.694 & 0.556 & 13.8 \\ 
SEE & 0.824 & 0.665 & 15.9 & 0.706 & 0.584 & 12.2 \\ 
MUSER & 0.841 & 0.693 & 14.8 & 0.729 & 0.617 & 11.2 \\ 
L-Defense & 0.773 & 0.671 & 10.2 & 0.695 & 0.608 & 8.7 \\ 
\cline{1-7}
{RoE-FND} & \textbf{0.860} & \textbf{0.843} & \textbf{1.7} & \textbf{0.743} & \textbf{0.712} & \textbf{3.1} \\
\hline
  \end{tabular}
  }
\label{tab:crossdatasets}
\end{table}

\subsection{Performance Comparisons}



\noindent\textbf{Comparisons on Text-only FND Datasets.}
We compare RoE-FND with multiple text-only baselines as shown in Table~\ref{tab:comparison}.
Including neural network methods and LLM-based methods. Details of them are presented in section~\ref{sec:related_works}.
We report accuracy (ACC), F1-macro (F1), precision (PR), and recall (RC), following previous works. 
RoE-FND consistently outperforms trained neural networks and other LLM-based approaches.
Several comparisons highlight specific advantages of RoE-FND.
RoE-FND and ProgramFC both decompose detection into sub-tasks, and the better performance with same models underscores the insight of our sub-task division.
Secondly, IMRRF integrates knowledge graph in addition to web searched evidence. 
Our better results highlight the benefits conferred by RoE-FND's experiential learning mechanism in addition to external evidence's help.
Moreover, we perform knowledge distillation from DeepSeek-R1 to Qwen-7B, denoted as RoE-FND$^*$ in the table.
Due to that the distillation data is generated with ground-truth posterior supervision, we regard the distilled variants as "trained" in our experimental reports.


\needmodify{
\noindent\textbf{Comparisons on Multimedia FND Datasets.}
In addition, we employ Qwen-VL in RoE-FND to handle multimedia samples. 
Results on FakeTT and FakeSV are reported in Table~\ref{tab:multimedia}.
The plain Qwen-VL model and the RAG approach achieve limited results on both datasets, 
highlighting the inherent difficulty of detecting fake news videos via logical deduction.
The other four methods learn deep learning features on the training set for detection, significantly outperforming Qwen-VL-based models. 
We see that RoE-FND achieves not only substantial gains over these four trained methods but also significantly surpasses the RAG method, suggesting its better reliability in logical deduction.
}


\begin{figure}[!t]
  \includegraphics[width=\columnwidth]{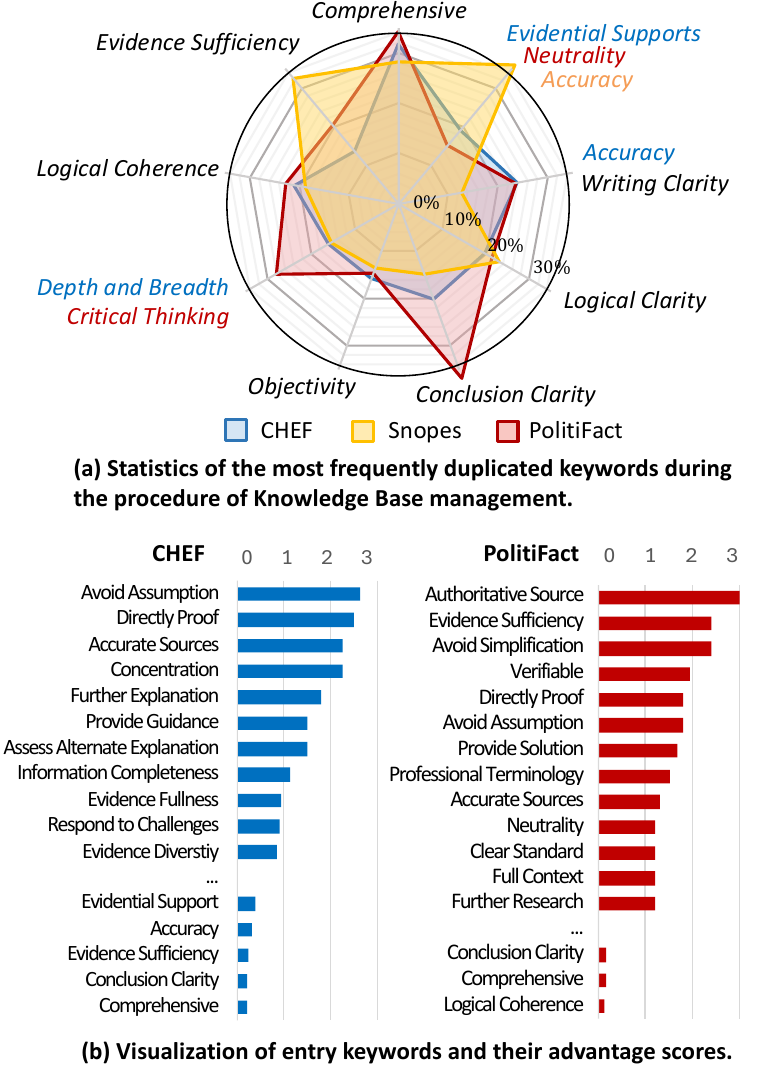}
  \caption{\ying{Experience-bank statistics across datasets. Keyword distributions indicate which verification criteria RoE-FND most often distills, offering an interpretable view of the framework’s acquired reasoning standards. Variations across datasets reflect domain-specific rumor characteristics, while consistently high-advantage criteria suggest transferable guidance that supports cross-dataset generalization.}}
  \label{fig:knowledge_base}
\end{figure}

\begin{figure*}[!t]
  \includegraphics[width=\linewidth]{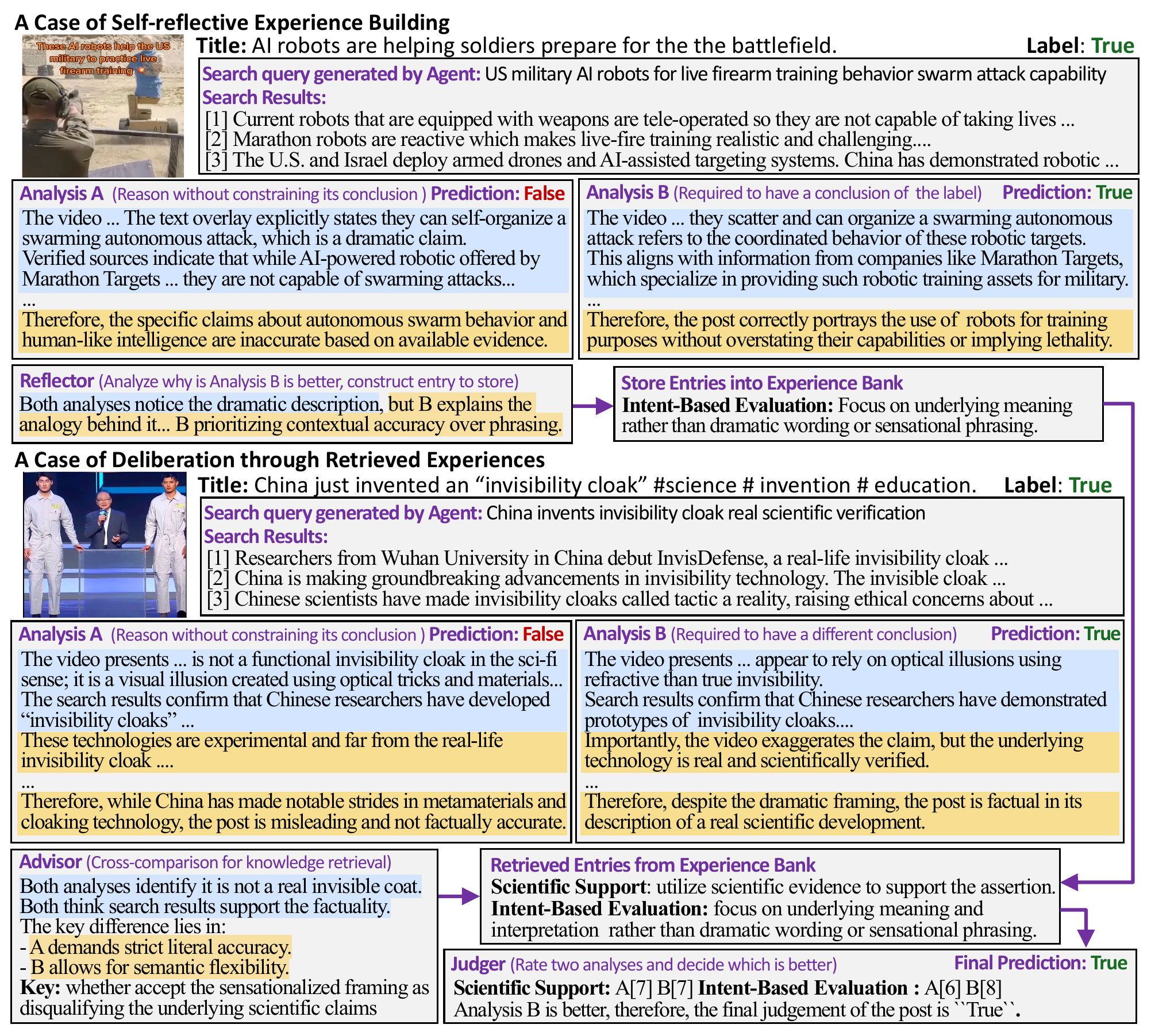}
  \caption{Two challenging cases of FakeTT from two stages of RoE-FND. We highlight analyses' consistency in light blue and inconsistency in light yellow. In the second stage, the learned experience helps the \texttt{Judger} make the correct prediction.}
  \label{fig:case_deploy}
\end{figure*}


\needmodify{\noindent\textbf{Cross-dataset \redmarker{Generalization} Performance.}
Apart from the above in-domain tests, we also conducted cross-dataset testing to evaluate the generalization ability of RoE-FND, which could be of higher value in application perspective.
The results are presented in Table~\ref{tab:crossdatasets}.
We select four baseline methods that require training. 
When evaluated on different datasets, these models exhibit a marked drop in performance. The results indicate that although the training process successfully teaches detection skills, it also inadvertently trains the models to be sensitive to the unique characteristics of the source dataset.
In contrast, RoE-FND demonstrates strong generalization even by constructing the experience bank on one dataset and testing on another. 
This ability likely stems from two factors: first, it detects through logical deduction rather than learning dataset-specific deep features. Second, the extracted experiences are not confined to a single data domain and can be transferred to unfamiliar data.}



\subsection{Experience Analysis and Case Study}

\noindent\textbf{Experience Bank Analysis.}
We perform experience bank management—comprising experience compression and refinement—on the training data. 
We count the most frequently duplicated keywords from the compression procedure, exhibit entries' keywords and their advantage scores. The results are visualized in Figure~\ref{fig:knowledge_base}. 
As will be testified in the ablation study, although the advantageous entries are discriminative during \texttt{Judger}'s assessment, they are not universally applicable.
One way to demonstrate this is that each dataset's advantageous entries adapt to and highlight its distinctive features.
For example, data from PolitiFact emphasizes concepts like ``neutrality'' and ``critical thinking,'' whereas rumors from Snopes place a greater premium on ``evidential support''.
However, as the results of cross-dataset testing show, such differences do not lead to evident performance dropping. Thus, we infer that the established experience bank already covers the necessary knowledge to handle the majority of cases.
Moreover, we find that entries with higher advantage scores tend to be more decisive for verification than those with lower scores. 
For instance, a keyword like ``avoid assumption'' provides greater discriminatory power between analyses than a term like ``comprehensive'' during \texttt{Judger}'s assessment procedure.
Notably, cross-comparing subfigures (a) and (b), the most frequently duplicated entries often receive low advantage scores. 
This demonstrates the necessity of experience management, by which these experiences are discarded to avoid hindering \texttt{Judger}'s reasoning.

\noindent\textbf{Case Study.}
\ying{In Figure~\ref{fig:case_deploy},} we present two related cases from FakeTT, drawn from the two stages of RoE-FND, respectively.
The presented cases further illustrate how RoE-FND improves reasoning reliability through experience-guided deliberation. 
In the first case, while Analysis A relies on surface-level wording, Analysis B focuses on verifying the underlying meaning, prioritizing intent over literal phrasing. 
This leads to the generation of an experience entry termed ``intent-based evaluation''.
In the second case, the framework encounters a similar divergence, this time over the interpretation of the term ``invisibility cloak''. \texttt{Advisor} identifies the disagreement and retrieves relevant entries from the experience bank, which includes the previously stored entry of ``intent-based evaluation''. 
Retrieved experiences help the \texttt{Judger} avoid overly literal or assumption-driven conclusions and instead prioritize semantically grounded verification principles.
Furthermore, the framework’s interpretable outputs make the model’s decision-making process far more transparent and accessible to human users.

\begin{table}[!t]
\centering
\caption{Ablation studies of the proposed RoE-FND. {$\downarrow$}(\%) indicates the decrease compared to the baseline setting.}
\setlength{\tabcolsep}{0.6mm}
{
  \begin{tabular}{l|rr|rr|rr}
  \hline
  \multirow{2}{*}{\centering\textbf{Ablation Setting}} & \multicolumn{2}{c|}{CHEF} & \multicolumn{2}{c|}{Snopes} & \multicolumn{2}{c}{PolitiFact}
  \\
  \cline{2-7}
     & {ACC}  & {$\pm$(\%)} & {ACC}  & {$\downarrow$}(\%) & {ACC} & {$\downarrow$}(\%)\\ 
     \hline
\textit{w/o} \texttt{Reflector} & 0.610 & -21.8 & 0.633 & -19.1 & 0.549 & -19.4 \\ 
\textit{w/o} \texttt{Advisor} & 0.748 & -8.0 & 0.721 & -10.3 & 0.555 & -18.8 \\ 
\textit{w/o} Dual Analysis & 0.725 & -10.3 & 0.736 & -8.8 & 0.645 & -9.8 \\ 
\textit{w/o} KB Management & 0.788 & -4.0 & 0.800 & -2.4 & 0.709 & -3.4 \\ 
Fixed Retrieval  & 0.704 & -12.4 & 0.776 & -4.8 & 0.678 & -6.5 \\ 
Randomly Retrieval  & 0.694 & -13.4 & 0.793 & -3.1 & 0.673 & -7.0 \\ 
Retrieval by News & 0.660 & -16.8 & 0.782 & -4.2 & 0.683 & -6.0 \\
\hline
\textbf{Baseline setting} & \textbf{0.828} & \textbf{-} & \textbf{0.824} & \textbf{-} & \textbf{0.743} & \textbf{-} \\ 
    \hline
  \end{tabular}
  }
\label{tab:ablation}
\end{table}

\subsection{Ablation Studies}
Table~\ref{tab:ablation} reports the ablation studies where we evaluate the necessity of model components and strategy of experience utilization.

\noindent\textbf{Ablations of Components.}
\noindent1)~\textit{w/o~\texttt{Reflector}} stores analyses from \texttt{Analyst} directly in the experience bank and presented to the \texttt{Judger} as exemplars. 
The performance drop indicates that unrefined, case-specific analyses are insufficient to be directly utilized to boost the reasoning ability of models.
\texttt{Reflector} successfully turns raw analyses into a solid and reusable experience.
\noindent2)~\textit{w/o~\texttt{Advisor}} fetches entries without cross-comparing two analyses. 
The decline suggests that the semantic information of analyses is not applicable for retrieving proper experience.
\noindent3)~\textit{w/o~\texttt{InformedAnalyst}} replaces the dual analysis with an \texttt{IntuitiveAnalyst}. 


\noindent\textbf{Experience Retrieval Strategy Variations.}
\noindent1)~\textit{w/o KB Management} disables experience compression and refinement, leaving redundant and low-quality entries in the bank, which increases retrieval noise and makes the Judger’s scoring less discriminative.
\noindent2)~\textit{Fixed retrieval} always selects a small set of globally high-advantage experiences. Although these experiences are generally useful, forcing them onto every case ignores the fact that different news items fail for different reasons. And 3)~\textit{Randomly retrieval} retrieves random experiences from the experience bank. 
It can provide irrelevant guidance that cannot help the \texttt{Judger} identify subtle reasoning flaws for deliberation.
\noindent3)~\textit{Retrieval by news} stores news-experience pairs in the bank and retrieves experiences according to the similarity between input news samples. Results indicate that news-level similarity does not necessarily imply reasoning-level similarity.

\section{Conclusions}

We propose RoE-FND, which introduces an experiential learning mechanism that distills reusable reasoning guidelines from reflective comparison and reuses them via retrieval to guide future decisions. Extensive experiments on benchmark datasets demonstrate strong effectiveness and cross-dataset generalization.


\begin{acks}
This work is supported by the National Natural Science Foundation of China 62572125, the Natural Science Foundation of Shanghai 25ZR1401019, and the China Postdoctoral Science Foundation under Grant 2025M771574.
\end{acks}

\appendix

\section{Additional Implementation Details}


\noindent\textbf{Embedding and Similarity Computation.}
We utilize sentence-level embedding created by all-MiniLM-L6-v2~\footnote{\url{https://huggingface.co/sentence-transformers/all-MiniLM-L6-v2}} for both retrieval and duplicate detection, and compute cosine similarity between normalized embeddings. The retrieval threshold is empirically set to $\tau_re = 0.9$, while the duplicate-filtering threshold is set to $\tau_s = 0.97$.

\noindent\textbf{Consistency Checking.}
If the prediction of \texttt{IntuitiveAnalyst} is correct, we further repeat inference two additional times to evaluate prediction consistency. Only samples that remain consistently correct are excluded from reflective experience generation. 

\noindent\textbf{LoRA Distillation Settings.}
We utilize standard causal language modeling loss for supervised fine-tuning. The LoRA rank is set to $r=16$, the scaling factor is set to $\alpha=32$, and the dropout rate is set to $0.05$. Training is conducted using the AdamW optimizer with a learning rate of $2\times10^{-5}$ and cosine learning rate scheduling.

\begin{table}[!t]
\centering
\caption{Average amount of output tokens of each RoE-FND component across different backbone models and datasets.}
\setlength{\tabcolsep}{0.8mm}
\renewcommand{\arraystretch}{1.15}
\begin{tabular}{llrrrrr}
\hline
 & \textbf{Dataset} & \textbf{Intuitive} & \textbf{Informed} & \textbf{Reflector} & \textbf{Advisor} & \textbf{Judger} \\ \hline

\multirow{3}{*}{\rotatebox{90}{\small DeepSeek}} 
& CHEF      & 341.4 & 302.8 & 124.5 & 77.3 & 42.5 \\ 
& Snopes    & 213.4 & 193.1 & 104.4 & 44.0 & 37.2 \\ 
& PolitiFact& 206.2 & 191.9 & 115.1 & 47.5 & 39.5 \\ \hline

\multirow{3}{*}{\rotatebox{90}{\small 4o-mini}} 
& CHEF      & 340.5 & 339.8 & 127.2 & 62.9 & 37.9 \\ 
& Snopes    & 262.0 & 240.4 & 109.0 & 42.4 & 50.2 \\ 
& PolitiFact& 251.1 & 222.6 & 123.5 & 41.2 & 52.3 \\ \hline

\multirow{2}{*}{\rotatebox{90}{\footnotesize QwenVL}} 
& FakeTT    & 234.5 & 206.8 & 121.2 & 50.1 & 40.8 \\ 
& FakeSV    & 165.0 & 158.3 & 111.0 & 48.6 & 48.2 \\ \hline

\end{tabular}
\end{table}

\begin{table}[!t]
\centering
\caption{Comparisons of the average amount of output tokens per sample during deployment for testing.}
\setlength{\tabcolsep}{1.2mm}
\begin{tabular}{lrrrr}
\hline
 \textbf{Method} & \textbf{Ref.} & \textbf{CHEF} & \textbf{Snopes} & \textbf{PolitiFact} \\ \hline
Self-Reflect & \cite{madaan2023selfreflect} & 532.4 & 350.9 & 363.8 \\
D2D & \cite{han2025debate} & 777.5 & 539.8 & 555.6 \\
DeepSeekR1 & \cite{guo2025deepseekr1} & 800.3 & 681.2 & 663.8 \\
RoE-FND & & 764.0 & 487.7 & 484.9 \\
\hline
\end{tabular}
\end{table}

\begin{figure}[!t]
    \centering
    \includegraphics[width=1.0\linewidth]{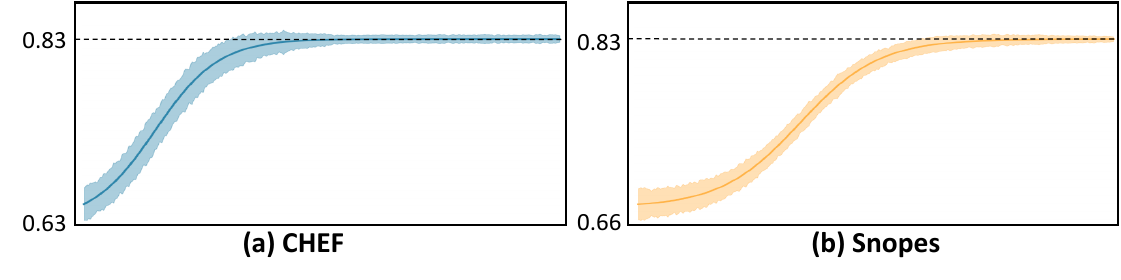}
    \caption{Learning curves of RoE-FND with respect to the size of the memory bank (from 0\% to 100\%).}
    \label{fig:placeholder}
\end{figure}

\section{Cost and Stability Analysis}

We report the average output token consumption to evaluate the practical inference cost.
The usage of output tokens directly reflects the computational and monetary costs incurred by each module.
The statistics show that most output tokens are consumed by the two \texttt{Analyst}s, while the \texttt{Advisor} and \texttt{Judger} introduce comparatively small overhead. 
The compact retrieved experiences help constrain the deliberation context length, resulting in better performance-cost tradeoff compared with other methods.

We visualize the growth of detection accuracy with the accumulation of the memory bank.
As the memory bank gradually grows, the performance of RoE-FND consistently improves on both datasets. This trend demonstrates that the proposed cross-sample review mechanism effectively benefits from accumulated historical cases. 
Meanwhile, the performance gain becomes saturated when the memory bank reaches a sufficient scale. This suggests that the framework captures most representative experiences after observing enough samples, and additional memory mainly provides marginal improvements. Such behavior indicates the favorable scalability of the proposed framework.

\section{Prompt of RoE-FND}

\includegraphics[width=1.0\linewidth]{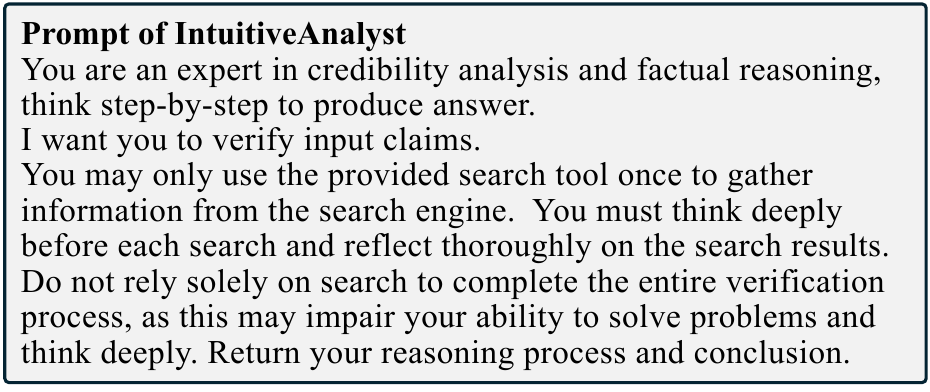}
\includegraphics[width=1.0\linewidth]{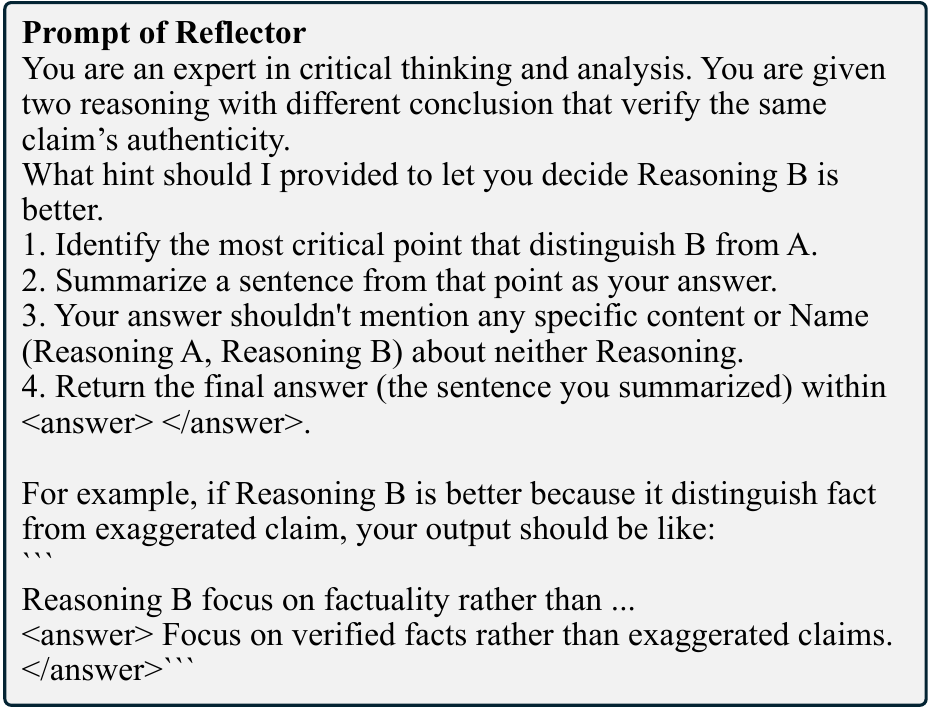}
\includegraphics[width=1.0\linewidth]{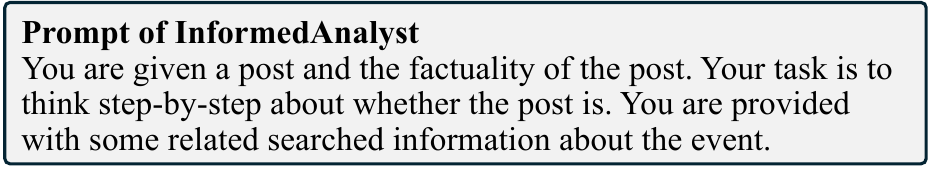}
\includegraphics[width=1.0\linewidth]{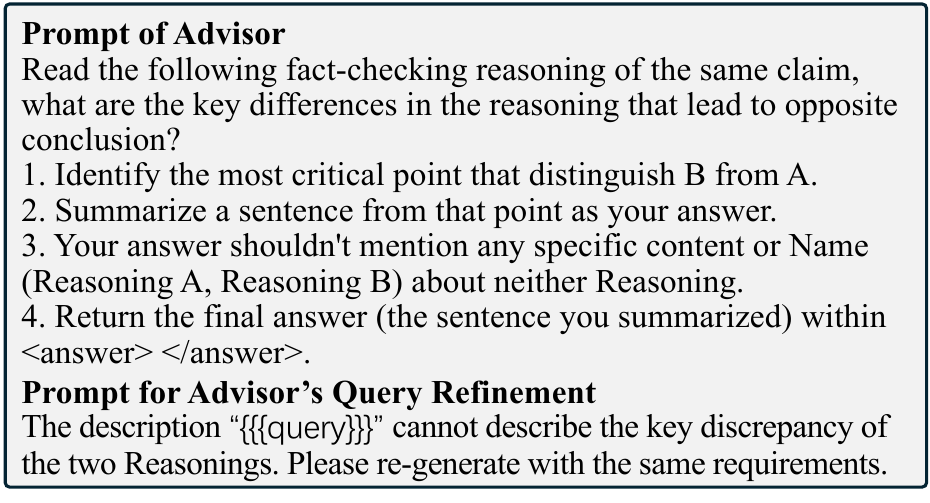}
\includegraphics[width=1.0\linewidth]{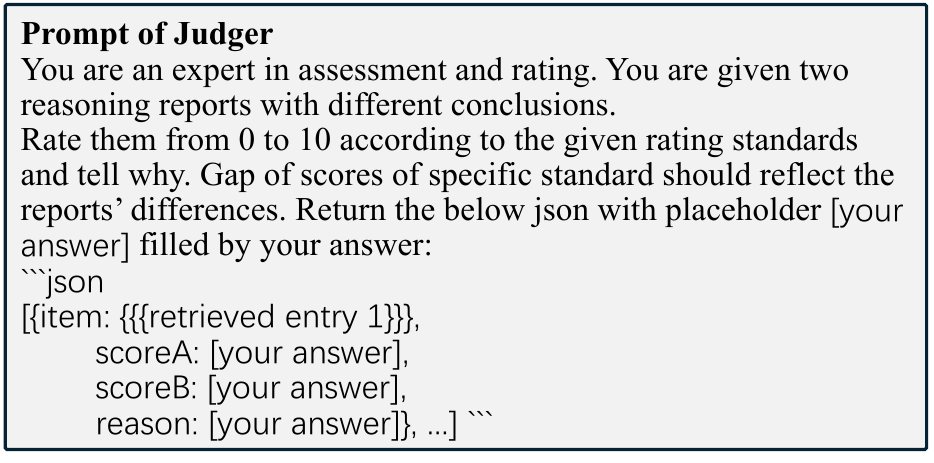}

\bibliographystyle{ACM-Reference-Format}
\balance
\bibliography{sigconf}

\end{document}